\documentstyle[epsfig,aps,amsmath,twocolumn,floats]{revtex}

\begin{document}
\draft
\wideabs{
\title{Hydrophobic interactions: an overview}

\author{Pieter Rein ten Wolde}
\address{FOM-Instiute for Atomic and Molecular Physics (AMOLF),
Kruislaan 407, 1098 SJ Amsterdam\footnote{Correspondence address}\\
Division of Physics and Astronomy, Vrije Universiteit, De Boelelaan 1081,
1081 HV, Amsterdam}

\maketitle
 
\begin{abstract}
We present an overview of the recent progress that has been made in
understanding the origin of hydrophobic interactions. We discuss the
different character of the solvation behavior of apolar solutes at
small and large length scales. We emphasize that the crossover in the
solvation behavior arises from a collective effect, which means that
implicit solvent models should be used with care. We then discuss a
recently developed explicit solvent model, in which the solvent is not
described at the atomic level, but rather at the level of a density
field. The model is based upon a lattice-gas model, which describes
density fluctuations in the solvent at large length scales, and a
Gaussian model, which describes density fluctuations at smaller length
scales. By integrating out the small length scale field, a Hamiltonian
is obtained, which is a function of the binary, large-length scale
field only. This makes it possible to simulate much larger systems
than hitherto possible as demonstrated by the application of the model
to the collapse of an ideal hydrophobic polymer. The results show that
the collapse is dominated by the dynamics of the solvent, in
particular the formation of a vapor bubble of critical
size. Implications of these findings to the understanding of pressure
denaturation of proteins are discussed.
\end{abstract}


}
\newpage

\section{Introduction}
 Hydrophobic interactions are widely believed to play a dominant role
in the formation of large biological
structures~\cite{Kauzmann59,Tanfordbook}. Yet, the mechanism of the
hydrophobic effect is still under debate. The solvation of small
apolar species is well
understood~\cite{Pratt77,Pangali79,Chandler93,Hummer96}. However, the
attraction between two such species in water is
weak~\cite{Pratt77,Pangali79,Hummer96} and probably not responsible
for the stability of biological structures. On the other hand, strong
and long-ranged attractions have been measured between extended
hydrophobic surfaces~\cite{Carambassis98,Tyrrell01}. But here, the
origin of the effect is still being discussed. It has been suggested that the
interaction arises from electrostatic fluctuations~\cite{Attard89},
changes in water structure~\cite{Eriksson89}, bridging
(sub)microscopic bubbles~\cite{Parker94,Attard00} and a ``drying''
transition induced by the hydrophobic
surfaces~\cite{Evans90,Attard92,Lum99}.

Recently, Lum, Chandler and Weeks~\cite{Lum99} have developed a theory
of hydrophobicity, which suggests that the nature of the hydrophobic
effect precisely arises from the interplay of density fluctuations at
both small and large length scales. Small apolar species only affect
density fluctuations in water at small length scales. Concomitantly,
water can only induce a relatively weak and short-ranged attraction
between small hydrophobic objects. In contrast, large hydrophobic
species can affect density fluctuations at large length scales. At
ambient conditions, water is close to phase coexistence. A
sufficiently large hydrophobic object, or more importantly, an
assembly of several small apolar species, can therefore induce a
depletion of water relative to the bulk
density~\cite{Stillinger73,Huang01_1}. Recently, it has been
demonstrated that this drying transition can induce a strong
attraction between hydrophobic objects and provide a strong driving
force for protein folding~\cite{TenWolde02_2}.

In this paper, we give an overview of the recent progress that has
been made in understanding the origin of the hydrophobic effect. In
section~\ref{sec:occup} we discuss the statistics of density
fluctuations at small and large length scales. Understanding these
fluctuations is important, because it does not only provide insight
into how the solvation free energy of solutes scales with their size,
but also which models are needed to describe their solvation
behavior. In the next section, we briefly discuss a recently developed
model. In section~\ref{sec:scaling}, we see that this model gives a
reasonable prediction of the solvation free energy of apolar
solutes. In particular, it predicts that in the small length scale
regime the solvation free energy scales with the size of the excluded
volume of the solute, whereas in the large length scale regime it
scales with the area of the excluded volume. The model also shows that the
crossover in the solvation free energy arises from a collective effect
in the solvent. Implicit solvent models cannot conveniently describe
this collective effect. It thus appears that
explicit solvent models are needed. Explicit atomistic solvent
models, however, are computationally demanding. The model discussed in
section~\ref{sec:theory}, lays the foundation for a scheme in
which the solvent is not described at the atomic level, but rather at
the level of a coarse-grained density field. This scheme, which is
discussed in section~\ref{sec:Langevin}, allows us to simulate the
solvent much more efficiently.

In section~\ref{sec:attractions} we discuss the role of attractive
 interactions between the solutes and the solvent. Most of the
 theoretical work on the hydrophobic effect has focused on the
 solvation behavior of ideal hydrophobic
 solutes~\cite{Lum99,Huang00_1,TenWolde02_1}. These are objects that
 exclude solvent from a certain region in space, but have no
 attractive interactions with the solvent. In
 section~\ref{sec:attractions}, however, it is seen that the presence of weak
 dispersive interactions does not significantly affect the solvation
 behavior of hydrophobic solutes.

Finally, in section~\ref{sec:polymer} we apply the scheme discussed in
section~\ref{sec:Langevin} to study the collapse of an ideal
hydrophobic polymer. The simulations reveal that the dynamics of the collapse
transition is dominated by the dynamics of the solvent. In particular,
the rate limiting step is the formation of a vapor bubble of critical
size. In addition, we show that during the collapse the chain and the
solvent remain out of equilibrium. Both observations imply that
implicit solvent models should be used with great care. In the past, a
statistical meaningful study of protein folding using explicit solvent
models seemed impractical. The current analysis, however, suggests
that such studies become feasible by using a statistical field model
like the one presented in sections~\ref{sec:theory}
and~\ref{sec:Langevin}.

\section{Density fluctuations at small and large length scales}
\label{sec:occup}
A theory of hydrophobicity should be able to correctly predict
the solvation free energy of apolar solutes. The solvation
free energy of a hard sphere, an example of an ideal hydrophobic
object, is given by
\begin{equation}
\Delta \mu(v_{\rm ex}) = -k_B T \ln[P(N=0;v_{\rm ex})].
\end{equation}
Here $k_B T$ is Boltzmann's constant times temperature and $P(N;v_{\rm
ex})$ is the probability of finding $N$ solvent particles inside a
region, of volume $v_{\rm ex}$, from which the solvent would be
excluded by the solute; note that $P(N=0;v_{\rm ex})$ corresponds to
the insertion probability of a hard sphere. Computer simulations have
revealed that density fluctuations that are entropic in origin, obey
Gaussian statistics. For the hard-sphere fluid, the distribution
$P(N;v_{\rm ex})$ (as a function of $N$) is found to be almost exactly
Gaussian for spherical volumes $v_{\rm ex}$ that have a diameter $d$
up to four times the
diameter of the solvent molecules, $\sigma$ (i.e. hard-sphere
diameter)~\cite{Crooks97}. Conceivably, the distribution functions
remain Gaussian for volumes that are larger than those that could be
studied in the simulations, albeit even larger voids could induce the
formation of a solid-fluid like interface if the fluid is close to
freezing. On the other hand, for volumes in SPC/E water at ambient
conditions~\cite{Hummer96} and a Lennard-Jones liquid close to phase
coexistence with the vapor~\cite{Huang00}, the distribution
$P(N;v_{\rm ex})$ remains Gaussian up to approximately
$d=1.0-2.0\sigma$. For such small volumes, the occupation statistics
is dominated by entropic effects. The solvation free energy of small
solutes is determined by the entropic cost of permitting only those
density fluctuations that do not violate the effect of the strong
forces that are present in the system. In a simple fluid, these strong
forces arise from the repulsive interactions between the particles; in
water, these are the hydrogen bond forces. As the occupation
statistics is Gaussian for small volumes, the insertion probability
$P(N=0;v_{\rm ex})$, and hence the solvation free energy of small
solutes, can be obtained from a Gaussian theory.  Examples of such
Gaussian theories are the Pratt-Chandler theory of
hydrophobicity~\cite{Pratt77,Chandler93} and the theory of Hummer,
Pratt and coworkers~\cite{Hummer96,Hummer98}.

 For larger volumes the situation changes significantly. In
a simple fluid, the attractive forces between the solvent molecules no
longer cancel each other when the volume $v_{\rm ex}$ is
excavated. The unbalanced attractive forces give rise to an
unbalancing potential, which tends to push the solvent molecules away
from the void. For water, a large excavated volume will drastically
disrupt the water structure. Close proximity of water to the void is
energetically unfavorable, because the hydrogen bond network can no
longer be maintained close to the surface of the void, as pointed out
by Stillinger in 1973~\cite{Stillinger73}. In both cases,
solvent density is depleted near the surface of the void. This drying
transition is a collective effect and can be interpreted as a
microscopic manifestation of a phase transition. A Gaussian model
cannot support such a transition as it only describes density
fluctuations at microscopic length scales. For larger solutes,
however, coarse-grained models, such as a Landau-Ginzburg model or a
lattice-gas model, become useful. Such coarse-grained models describe
density fluctuations at length scales larger than the bulk correlation
length and can support phase transitions and sustain gas-liquid
interfaces. The essence of our model, discussed in
section~\ref{sec:theory}, is to combine a Gaussian model with a
lattice-gas model.

\section{Theory of solvation}
\label{sec:theory}
Water at ambient conditions is a cold liquid. A cold liquid is a fluid
that is well below the critical temperature. For such a cold liquid,
the density of the vapor is typically orders of magnitude lower than
that of the liquid. It is then natural to divide space into a (cubic)
grid of cells and, taking the grid spacing,
$l$, to be on the order of the bulk correlation length, $\xi$, describe
the states of the cells  by a binary field, $n_i$,
where $n_i=1$ if cell $i$ is filled with ``liquid'' and $n_i=0$ if
cell $i$ contains ``vapor''~\cite{TenWolde02_1}. This binary field describes density
fluctuations at length scales larger than the grid spacing and it can
support phase transitions. Density fluctuations at length scales
smaller than the grid spacing are described by a second field, $\delta
\rho$. This field does not support phase transitions, but it can
describe the microscopic granularity of matter. As discussed above, we
can assume that this field obeys Gaussian statistics.  Hence, the full
density $\rho ({\bf r}_i)$ at point ${\bf r}_i$ is decomposed as:
\begin{equation}
\rho({\bf r}_i) = \rho_l n_i + \delta \rho({\bf r}_i),
\end{equation}
where $\rho_l$ is the bulk liquid density.

The partition function for
our model is
\begin{eqnarray}
\label{eq:Xi}
\Xi &=&\sum_{\{n_{i}\}}\int {\cal D} \delta \rho ({\bf r}) \: C[\left\{
n_{k}\right\},\delta \rho ({\bf r})] \: \nonumber \\
&&\times \: \exp \left(-\beta H[\left\{
n_{k}\right\} ,\delta \rho ({\bf r})]\right),
\end{eqnarray}
where $\int {\cal D}\delta \rho ({\bf r})=\int \Pi _{i}{\cal D}\delta \rho (
{\bf r}_{i})$ denotes the functional integration over the small length scale
field, $H[\left\{ n_{k}\right\} ,\delta \rho ({\bf r})]$ is the Hamiltonian
as a functional of both $n_{i}$ and $\delta \rho ({\bf r}),$ and $\beta
^{-1} $ is Boltzmann's constant times temperature, $k_{B}T$. The
quantity $C[\left\{ n_{k}\right\} ,\delta \rho ({\bf r})]$ is a constraint
functional. It has unit weight when the field $\delta \rho ({\bf r})$ together with
$\left\{ n_{i}\right\} $ satisfy whatever constraints are imposed by strong
forces, and it is zero otherwise. Since $\{n_i\}$ and $\delta
\rho({\bf r})$ have greatly different character, the summation and
integration in Eq.~(\ref{eq:Xi}) do not redundantly count configuration
space to any significant degree.

The Hamiltonian of our model is given by the following expression:
\begin{eqnarray}
 H[\left\{ n_{k}\right\},\delta \rho ({\bf r})] & = &
 H_{L}[\left\{
n_{k}\right\} ]-\epsilon^{\prime }\negthickspace \sum_{i,j ({\rm nn}i)}\int d{\bf r}_{i} \,
\delta \rho({\bf r}_{i}) \, \frac{n_{j}-1}{\rho _{l}l^{3}} \nonumber \\[0.10cm] &&+ \: \frac{k_{B} T}{2}\sum_{i,j}\int d{\bf r}_{i}\int d{\bf
r}_{j}^{\prime} \:\nonumber \\[0.10cm]
&&\hspace*{1.0cm}
\times \: \delta \rho ({\bf r}_{i}) \: \chi ^{-1}[{\bf
r}_{i},{\bf r}_{j}^{\prime};\{n_{k}\}] \: \delta \rho ({\bf
r}_{j}^{\prime }) \nonumber \\[0.10cm]
&&+ \: H_{\mbox{\tiny norm}}[\left\{ n_{k}\right\}].  \label{eq:H}
\end{eqnarray}

In the above expression, $H_L$ is the Hamiltonian of a lattice-gas
model. It is given by
\begin{equation}
\label{eq:H_L}
H_{L}[\left\{ n_{k}\right\} ]=-\mu \sum_{i}n_{i}-\epsilon
\sum_{<i,j>}n_{i}n_{j},
\end{equation}
with $\mu$ the imposed chemical potential, and $\epsilon$ the
interaction parameter between nearest-neighbour cells. The lattice-gas model
describes density fluctuations at length scales larger than the
grid spacing and it can support phase
transitions. Importantly, the
energetic cost of creating an interface is determined by the value of
$\epsilon$. 

The term in Eq.~\ref{eq:H} that is quadratic in $\delta \rho$ ensures
the Gaussian weight of the small length scale field. In particular,
the response function $ \chi ^{-1}[{\bf r}_{i},{\bf
r}_{j}^{\prime};\{n_{k}\}]$ determines the variance of this field. It
depends upon the state of the large length scale field.  If $n_i=1$
for all $i$, then $\chi ^{-1}[{\bf r}_{i},{\bf
r}_{j}^{\prime};\{n_{k}\}]$ reduces to the response function of the
bulk liquid $\chi ^{-1}[{\bf r}_{i},{\bf r}_{j}^{\prime};\rho_l]$. In
general, we employ the approximation that $\delta \rho({\bf r}_i)=0$
whenever $n_i=0$. Thus, $\delta \rho({\bf r})$ is a Gaussian field,
with a weight functional being that of the bulk liquid, but constrained
to be zero whenever $n_i=0$~\cite{TenWolde02_1}.

The second term on the right-hand-side of Eq.~\ref{eq:H} is the term
that couples the fluctuations of the field $n_i$ with those of the
field $\delta \rho({\bf r})$. As discussed by Lum {\em et al.} and
Weeks and coworkers~\cite{Lum99,Weeks98}, for simple fluids the
parameter $\epsilon^\prime$ is related to the energy density of the
bulk liquid. The last term on the right-hand-side of Eq.~\ref{eq:H},
$H_{\rm norm}$, is a renormalization term, which ensures that, if the
field $\delta \rho$ is integrated out in the {\em absence} of solutes,
the Hamiltonian $H[\{n_k\};\delta \rho]$ reduces to that of the
lattice-gas model, $H_L[\{n_k\}]$.

In the {\em presence} of a hard sphere, the
constraint functional in Eq.~\ref{eq:Xi} is given by 
\begin{equation}
C[\{n_k\};\delta \rho({\bf r})] = \prod_{{\bf r}_{i}\in
v_{\mbox{\tiny ex}}}\delta [n_{i}\rho _{l}+\delta \rho ({\bf r}_{i})].
\end{equation}
By exploiting the Fourier representation of delta functions, we can
integrate out the field $\delta \rho({\bf r})$. As discussed in detail
in Ref.~\cite{TenWolde02_1}, we then arrive at the following
expression for the Hamiltonian:
\begin{eqnarray}
H[\left\{ n_{k}\right\} ] &=&H_{L}[\left\{ n_{k}\right\} ]
\nonumber \\
&&+ \: k_{B}T \sum_{i,j({\rm occ})}\frac{n_{i}\left[ \rho _{l}v_{i}+f_{i}\right] \left[
\rho _{l}v_{j}+f_{j}\right] n_{j}}{2\sigma _{v_{\mbox{\tiny
ex}}}}\nonumber\\
&&+ \: k_{B}T\ln \sqrt{2\pi \sigma _{v_{\mbox{\tiny ex}}}};  \nonumber \\[0.10cm]
&\equiv &H_{L}[\left\{ n_{k}\right\} ]+H_{S}[v_{\mbox{\tiny ex}
};\{n_{k}\}].  \label{eq:H_fs}
\end{eqnarray}
Here, 
\begin{equation}
f_{i}\equiv \int_{{\bf r}_{i}\in v_{\mbox{\tiny ex}}} \negthickspace d{\bf r}_{i}f(
{\bf r}_{i})= n_{i}\:v_{i}\:\epsilon^{\prime }\:\kappa \:\frac{\rho _{l}}{l^{3}}\sum_{k({\rm nn}i)}(n_{k}-1),
\end{equation}
with $\kappa$  the isothermal compressibility, and
\begin{equation}
\sigma _{v_{\mbox{\tiny ex}}}=\int_{v_{\mbox{\tiny ex}}}d{\bf r}\int_{v_{
\mbox{\tiny ex}}}d{\bf r}^{\prime }\chi ({\bf r},{\bf r}^{\prime };\rho
_{l}).
\end{equation}
The sum over $i,j({\rm occ})$ is over cells $i$ and $j$ that are
occupied by the solute and $v_{i}$ is the volume occupied by the solute in cell
$i$~\cite{TenWolde02_1}.

The term $H_S[v_{\rm ex};\{n_k\}]$ contains all the effects of the
interaction between the ideal hydrophobic solute and the solvent. We
stress that the interaction term solely arises from the constraint
that is imposed upon the allowed density fluctuations in the
solvent. The excess chemical potential of the solute may be obtained
by averaging this interaction free energy:
\begin{equation}
\label{eq:bdmu}
\beta \Delta \mu (v_{\rm ex}) = -\ln\langle \exp(-\beta H_S[v_{\rm
ex};\{n_k\}]\rangle_L,
\end{equation}
where $\langle ... \rangle_L$ denotes an ensemble average with the
Hamiltonian of the lattice gas, $H_L[\{n_k\}]$ (see Eq.~\ref{eq:H_L}).

\section{Scaling behavior of solvation free energies}
\label{sec:scaling}
\begin{figure}[t]
\begin{center}
\epsfig{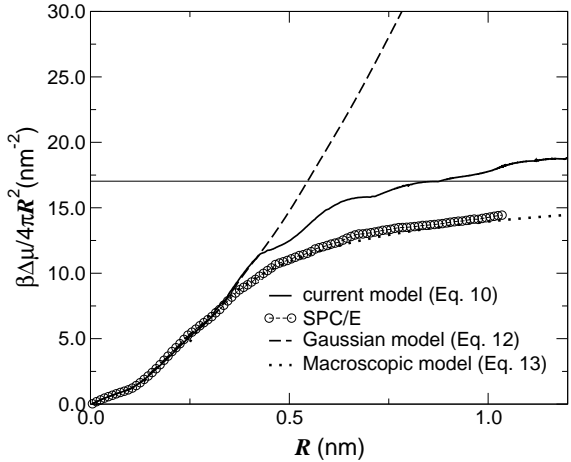}
\end{center}
\caption{\label{fig:bdmu_SPCE}Comparison of the results of the full
model, based on Eq.~\ref{eq:H_fs} and Eq.~\ref{eq:bdmu}, with the
predictions of the Gaussian model, Eq.~\ref{eq:bdmu_G}, and the
results of a molecular simulation of a cavity in SPC/E
water~\protect\cite{Huang01_1} for the excess chemical potential per unit area
of a hard sphere in water as a function of its size; the hard sphere
excludes water from a spherical volume of radius $R$.  The horizontal
line lies at the value of the surface tension $\gamma$ of the
vapor-liquid interface of water. The parameters of the full model are:
$l=4.2$\AA~, $\epsilon=6.02k_BT$, $\epsilon^{\prime}=15.2k_BT$, and
$\mu-\mu_{\rm coex}=5.5 \times 10^{-4} k_B T$. These parameters are
chosen such that the surface-tension, the energy density and the
imposed chemical potential of the lattice-gas model matches the
corresponding values of water (see Ref.~{\protect\cite{TenWolde02_1}}).}
\end{figure}

Fig.~\ref{fig:bdmu_SPCE} shows the excess chemical potential of a hard
sphere as a function of its radius, $R$, in water at ambient
conditions, as predicted by Eqs.~\ref{eq:H_fs} and~\ref{eq:bdmu}. For
comparison, we also show the results of a Gaussian model and the
results of a molecular simulation of a hard sphere in SPC/E
water~\cite{Huang00}.

\subsection{Small length scale regime}
 In the small length scale regime, our model reduces to a
Gaussian model. Small solutes only affect density
fluctuations in the solvent at small length scales. As a consequence,
$n_i=1$ for nearly all $i$. With $n_i=1$ for all $i$, $H_L[\{n_k\}]$
and $H_{\rm norm}[\{n_k\}]$ in Eq.~\ref{eq:H} become constants and
thus irrelevant. Further, the response function reduces to that of the
uniform fluid and the term that couples the density fluctuations of
the small and large length scale field, becomes zero. As such, the
Hamiltonian reduces to that of a Gaussian model~\cite{Chandler93}:
\begin{equation}
H_{G}\left[ \delta \rho ({\bf r})\right] =\frac{k_{B}T}{2}\int d
{\bf r}\int d{\bf r^{\prime }}\delta \rho ({\bf r})\chi ^{-1}({\bf r},{\bf
r^{\prime }};\rho _{l})\delta \rho ({\bf r^{\prime }}),  \label{eq:H_G}
\end{equation}
with $\delta \rho ({\bf r})=\rho ({\bf r})-\rho _{l}$, and the
response function $\chi ^{-1}({\bf r},{\bf r^{\prime }};\rho _{l})$
being the response function of the uniform fluid. The Pratt-Chandler
theory of hydrophobicity is based upon such a Gaussian
model~\cite{Pratt77}. Similarly, we can directly obtain the solvation free
energy of a Gaussian model  from
Eq.~\ref{eq:bdmu}. It is given by:
\begin{equation}
\beta \Delta \mu(v_{\rm ex}) \simeq \rho_l^2 v_{\mbox{\tiny ex}
}^{2}/2\sigma _{v_{\mbox{\tiny ex}}}+\ln \sqrt{2\pi \sigma _{v_{
\mbox{\tiny ex}}}}.  \label{eq:bdmu_G}
\end{equation}
This is also the result for the solvation free energy of the
theory of Hummer, Pratt and coworkers~\cite{Hummer96,Hummer98}.

Fig.~\ref{fig:bdmu_SPCE} shows that
 the full model and (thus) the Gaussian model give an excellent prediction for
the solvation free energy of small solutes. The reason, as discussed
in more detail in section~\ref{sec:occup}, is that at small length
scales, density fluctuations are highly Gaussian. A second point to
note is that for small solutes, the excess chemical potential scales with the
excluded volume of the solute. In this regime, the excess chemical
potential is dominated by the entropic cost of constraining small
length scale fluctuations, which is proportional to the volume of the
space over which the solvent density fluctuations are constrained.

\subsection{Large length scale regime}
For solutes larger than $R>4$\AA, the prediction of the Gaussian model
diverges from both the full model and the molecular simulation
results. The divergence is due to a cavitation transition in the
solvent. This cavitation transition can be understood as a microscopic
liquid-gas phase transition in the solvent that is induced by the
solute. In order to describe the solvation behavior of large
hydrophobic objects in this ``drying'' regime, a theory of
solvation should be able to support a liquid-gas phase transition and
capture the enthalpic effects of creating a gas-liquid interface. The
full model discussed in the previous section is such a model. A
Gaussian model, on the other hand, cannot support a gas-liquid phase
transition, nor sustain a gas-liquid interface, as it is based upon a
density expansion around the uniform fluid.

Fig.~\ref{fig:bdmu_SPCE} shows that in the large length scale regime,
the molecular simulation results for the excess potential are well
described by~\cite{Huang01_1}
\begin{equation}
\label{eq:dmu_therm}
\frac{\Delta \mu(R)}{4\pi R^2} \approx \frac{p R}{3} + \tilde{\gamma}
(1-\frac{2 \delta}{R}).
\end{equation}
The first term on the right-hand side is the work to expand a cavity
against the external pressure, $p$. The second term on the right-hand
side of the above equation describes the work to form an interface
between the solute and the solvent, including a first-order ``Tolman''
correction due to the curvature of the interface.

 At ambient conditions, the
pressure is very small, and the $p V$ contribution to the free energy is
only significant for extremely large solute volumes. Moreover, as
discussed in detail in Ref.~\cite{Huang01_1}, $\tilde{\gamma}$ is well
approximated by the surface tension of the vapor-liquid
interface at coexistence. Indeed, for large (but not
unphysically large) solutes, the solvation free energy is dominated by
the work to create a gas-liquid interface.

It is often assumed that the solvation free energy of a hydrophobic
species is proportional to the exposed surface
area~\cite{Eisenberg86,Vallone98}. The molecular simulations, as well
as the theoretical analyses, indicate that for solutes of biologically
relevant size, this is a reasonable assumption in the drying
regime. For small solutes, however, the excess chemical potential of a
hydrophobic species does not scale with the area of the interface
between the solute and the solvent. Rather, in the small length scale
regime, the excess chemical potential scales with the excluded volume
of the solute. Finally, we point out that while macroscopic
thermodynamic descriptions such as that underlying
Eq.~\ref{eq:dmu_therm} are useful for understanding the solvation
behavior of large solutes ($R>8-10$~\AA), they are of limited use for
describing the solvation behavior of small solutes. In the small
length scale regime, microscopic models, such as the Gaussian model
discussed in the previous section, are required.

\section{Langevin dynamics in Ising solvent}
\label{sec:Langevin}
The model discussed in section~\ref{sec:theory} can be used to develop
a scheme in which only the solutes are treated at the atomic level;
the solvent is described in terms of the binary, large-length scale
density field $n_i$. In this way, the solvent can be simulated much
more efficiently than using an explicit atomistic solvent model. The
scheme is generally applicable to describe the motion of a collection
of solutes, but here we will confine our attention to the motion of an
ideal hydrophobic polymer consisting of $N_s$ hard spheres.

Before we discuss the scheme in detail, we point out that, in
practice, it is reasonable to simplify the Hamiltonian in
Eq.~\ref{eq:H} by neglecting the relatively small off-diagonal
elements~\cite{TenWolde02_2}. This yields the following Hamiltonian for the coarse-grained
density field:
\begin{equation}
\label{eq:V_Ss}
H({\bf r}^{N_s};\{n_k\}) = \sum_i \left (-\mu + \Delta \mu_{\rm
ex} (v_i)\right ) n_i -
\epsilon \sum_{<i,j>} n_i n_j.
\end{equation}
In the above expression, $v_i= \sum_{s=1}^{N_s} v_{s,i}$, where
$v_{s,i}$ is the volume that monomer $s$ occupies in cell $i$ and
$N_s$ is indeed the number of monomers with coordinates ${\bf
r}^{N_s}$. Further, $\Delta \mu_{\rm ex} (v_i)$ is the reversible
work to accommodate a volume $v_i$ in the solvent. We have taken
$\Delta \mu_{\rm ex}$ to be proportional to the excluded volume:
\begin{equation}
\label{eq:dmu_v}
\Delta \mu_{\rm ex} (v_i) \approx c v_i,
\end{equation}
 with $c=2.67\times 10^8$J/m$^3 = 65k_B T/$nm$^3$ at room temperature.

From $H({\bf r}^{N_s};\{n_k\})$, we can construct a free-energy
functional, $\Omega$, for the field $\{n_k\}$ in the mean-field
approximation:
\begin{multline}
\label{eq:Omega}
\Omega ({\bf r}^{N_s};\{n_k\})=\\
 \sum_i k_B T (\langle
 n_i \rangle \ln \langle n_i \rangle  + (1-\langle n_i
\rangle) \ln (1-\langle n_i \rangle))\\
+\sum_i \left(-\mu + \Delta
\mu_{\rm ex} (v_i)\right) \langle n_i \rangle -\frac{\epsilon}{2}
 \sum_{<i,j>}\langle n_i\rangle \langle n_j \rangle 
\end{multline}
We can now construct a Car-Parinello
scheme~\cite{Car85,Frenkelbook} to propagate the solvent. To this end, we
define the Lagrangian:
\begin{eqnarray}
{\mathcal L}({\bf r}^{N_s};\{n_k\}) &=& \frac{1}{2}
 \sum_{\alpha=1}^{N_s} M \dot{{\bf r}}_\alpha^2 - V_{SS}({\bf
 r}^{N_s})\nonumber\\
&& + \frac{1}{2}\sum_i
 m \langle \dot{n}_i\rangle^2 - \Omega ({\bf r}^{N_s};\{ n_k \}).
\end{eqnarray}
Here $V_{SS}({\bf r}^{N_s})$ is the intra-chain potential that describes the
  direction interaction between the monomers, $M$ is the mass of the monomers and $m$ is a fictitious
  mass associated with the dynamical variable $\langle n_i
  \rangle$. From the Lagrangian, the following equation-of-motions for
  the monomers and the variables $\langle n_i \rangle$ is obtained:
\begin{eqnarray}
M \ddot{{\bf r}}_{\alpha} &=&-\nabla_{\alpha }{\mathcal L} =
 -\nabla_{\alpha }[V_{SS}({\bf r}^{N_s}) + \Omega({\bf
 r}^{N_s};\{n_k\})]\\
 m \langle \ddot{n}_i \rangle &\equiv&
\frac{\partial{\mathcal L}}{\partial \langle n_i \rangle} =-
\frac{\partial \Omega ({\bf r}^{N_s};\{n_k\})}{\partial \langle n_i
\rangle}.
\end{eqnarray}
Propagating the solvent field $\{n_k\}$ according to the above
equation-of-motions with a proper choice of both the fictitious mass
$m$ and the kinetic energy associated with $\{\langle n_k \rangle \}$,
ensures that the free energy of the solvent is close to its minimum
for each configuration of the polymer~\cite{Frenkelbook}. This
approach is similar in spirit to the method employed by L\"{o}wen,
Madden and Hansen to simulate counterion screening in colloidal
suspensions of polyelectrolytes~\cite{Lowen92}. However, as we will
demonstrate below, we cannot assume that the chain moves slowly on
time scales for which disturbances in the solvent may relax. As we
will see, the collapse of the polymer is driven by a cavitation
transition in the solvent. Such a rare event cannot be captured by the
above Car-Parinello scheme as nucleation barriers cannot be
crossed. This means that, using a Car-Parinello scheme, the solvent would
remain in its metastable state until it would become unstable. We therefore
developed a novel scheme in which the dynamics of the solvent and the chain
are treated together.

First,  it should be realized that the
small length scale field, $\delta \rho$, has been integrated out. This
means that trajectories are only true to nature on time scales larger
than those required to relax the small length scale field $\delta
\rho$. This relaxation time is on the order of
picoseconds~\cite{Stillinger75}, which implies that the dynamics is
diffusive. We thus constructed a stochastic dynamics scheme. The
elementary step of the algorithm consists of propagating the polymer
for $M_S$ Langevin steps in the field of constant $\{n_k\}$, followed
by a full Glauber sweep~\cite{Landau00} over the solvent field $\{n_k\}$. One Langevin
step corresponds to:
\begin{multline}
{\bf r}_{\alpha }(t+\delta t_s) = {\bf r}_{\alpha }(t)\\
 +
\frac{\delta  t_s }{\gamma}(-\nabla _{\alpha }[ V_{SS} ({\bf
 r}^{N_s}) + H  ({\bf r}^{N_s};\{n_k\})] + \delta {\bf F}).
\end{multline}
where $\gamma$ is a friction coefficient, the value of which can be
obtained from the diffusion constant of a single monomer in water. The
beads also experience a random force $\delta {\bf F}$; it is the dynamical
remnant of the small length-scale field. In order to obtain a
physically meaningful value for $M_s$, we have to compare the time
step for the propagation of the polymer, $\delta t_s$, to the time
scale that corresponds to a Glauber sweep over the solvent
variables. An estimate for the latter can be obtained by estimating
the correlation time for a density fluctuation of length scale $l$:
$\delta t_l=1/[D(2\pi/l)^2]$, where $D$ is the self-diffusion constant
of liquid water. The value of $M_s$ is thus given by $M_s = \delta
t_l/\delta t_s = 36$.

\section{Attractions}
\label{sec:attractions}
So far, we have discussed the solvation of ideal hydrophobic solutes
-- objects that have no attractive interactions with the solvent
molecules. Naturally, real hydrophobic molecules have some affinity
for water, and {\em vice versa}, because of the ubiquitous Van der
Waals interactions. Huang and Chandler have studied in detail the
effect of weak solute-solvent attractions on the solvation of
non-polar molecules in water~\cite{Huang02}. Their analysis was
performed using an extension and improved parameterization of the
theory of Lum, Chandler and Weeks~\cite{Lum99}.

Huang and Chandler showed that for small solutes, the presence of
attractions has little effect on the solvent density around the
solutes, as has previously been appreciated
theoretically~\cite{Pratt77} and as observed in
simulations~\cite{Guillot93}. The solvation behavior can be understood
from the observation that the attractive forces are very small in
comparison with hydrogen bond forces, and their effects can be
estimated by assuming that the water structure around a hard sphere
is unaltered by adding an attraction to water. In particular,
attractive interactions should produce a simple additive contribution
to the solvation enthalpy, but no significant effect on the solvation
entropy. 

For large hydrophobic species, attractions do have a notable
effect. In the case of hard spheres, ie. without attractions, the
solvent density near the surface of the solute is strongly depleted
relative to that in the bulk liquid. In contrast, in the presence of
attractions, the solvent density near the surface of the
solute is close to that in the bulk liquid ($g(R^+)\approx 1.24$, where
$g(r)$ is the solvent radial distribution function and $R$ is the
radius of the hydrophobic sphere). This, however, is a result of the
fact that the drying layer is very compressible. In particular, there
is little free-energy cost to move the vapor-liquid interface. Due to
this small energetic cost, the addition of an attractive potential as
weak as that between alkanes and water is sufficient to draw the
drying layer into contact with the hydrophobic surface.

While drawn into contact with the hydrophobic surface, this interface
is distinct from the interface that surrounds a small hydrophobic
solute. The contact values for $g(r)$ for the large hydrophobic
solutes are close to one, while the small hydrophobic solutes have
contact values larger than two. More importantly, as for ideal
hydrophobic objects, the solvation free energy scales with the size of
the excluded volume for smaller solutes, whereas the solvation free
energy scales with the area of the excluded volume for larger
solutes. In addition, the crossover from the small to the large length
scale regime is around $8-10$\AA~ for both the ideal hydrophobic
objects and the hydrophobic solutes that have attractive interactions
with the solvent. Thus, the scaling behavior of the solvation free
energy is not affected by the presence of weak attractions between the
solute and the solvent. We therefore expect that weak dispersion
forces will not have a strong effect on the role of hydrophobic forces
in biological self-assembly.

\section{Hydrophobic polymer collapse}
\label{sec:polymer}
Hydrophobic interactions have long been considered to play an
important role in protein folding. We have performed computer
simulations to study the collapse of a polymer consisting of $N_s=12$
hard spheres in water. Even though this allows us to study the effect
of hydrophobic interactions on protein folding in perhaps its most
basic form, the conventional approaches will not work. The theoretical
analyses as discussed above, suggest that the hydrophobic effect
arises from a collective effect in the solvent. Implicit solvent
models cannot conveniently capture this effect. It thus appears that
explicit solvent models are required. Explicit solvent models,
however, are computationally demanding. The analysis is further
complicated by the fact that the crossover from solvation in the small
length scale regime to solvation in the large length scale regime
occurs at around a nanometer. This implies that in order to study the
hydrophobic effect, large system sizes are required. For the required
system sizes, an explicit atomistic model of water would not be
tractable in the simulations. In contrast, the scheme discussed in
section~\ref{sec:Langevin} makes it possible to study the polymer
collapse in water.

\begin{figure}[h]
\begin{center}
\epsfig{figure = 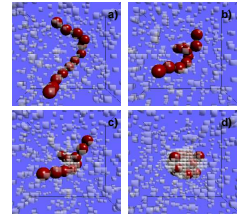,width=3.0cm}
\end{center}
\caption{\label{fig:snapshots}
Four configurations from a trajectory where a 12 unit hydrophobic
chain in water goes from the coil to the globule state. (a) shows a
configuration from the equilibrated coil. The chain remained in
configurations like that throughout a $10$~ns run at room temperature
($ T=0.663\epsilon $). On a much longer time scale, about $10^{-5}$~s,
the chain typically does exhibit a transition from coil to
globule. Such events are found with transition path sampling,
equilibrating from an initial high temperature ($T=0.74\epsilon$)
10~ns trajectory that exhibited the transition spontaneously. Three
configurations from an equilibrated $1.5$~ns trajectory that
exhibits the collapse transition at room temperature, are shown in
(b), (c) and (d), with that in (c) being a configuration from the
transition state surface.  The transparent cubes denote the vapor
cells. Those seen far from the chain are typical spontaneous density
fluctuations in bulk liquid water. The size of the simulation box is
$397~{\rm nm}^{3}$, corresponding to 42,875 cells.  The potential
energy function is given by $V_{SS}({\bf r}^{N_s}) + H ({\bf
r}^{N_s};\{n_k\})$. The latter term describes the solvent and the
interaction between the polymer and the solvent; it is given by
Eq.~\ref{eq:V_Ss}. The parameters are $l=0.21~$nm, $\mu-\mu_{\rm coex}
= 2.25 \times 10^{-4} k_B T$, and $\epsilon=1.51k_B T$. The
intra-chain potential, $V_{SS}({\bf r}^{N_s})$, is a function of the
positions of the centers of each of the 12 hydrophobic spheres (the
red particles in the figure). It contains three parts: (1) steep
(essentially hard sphere) repulsions between solute particles such
that their interparticle separations are larger than $\sigma
=0.72~$nm; (2) stiff harmonic potentials bonding adjacent particles in
the hydrophobic chain, $\frac{1}{2} k_{s}(\sigma -\left| {\bf
r}_{\alpha +1}-{\bf r}_{\alpha }\right| )^{2}$, with
$k_{s}=14.1$J/m$^{2}$; (3) a bending potential favoring an extended
chain, $\frac{1}{2}k_{\theta }\theta _{\alpha }{}^{2},$ where $\theta
_{\alpha }$ is the angle between $\left( {\bf r}_{\alpha +2}-{\bf
r}_{\alpha +1}\right) $ and $\left( {\bf r}_{\alpha +1}-{\bf
r}_{\alpha }\right) $, and $k_{\theta }=1.85\times
10^{-20}$J/rad$^{2}$. The volumes $v_{i}$ excluded from water by the
chain are dynamic as they change with changing chain configuration,
i.e., $v_{i}=v_{i}\left( \left\{ {\bf r}_{\alpha }\right\} \right)
$. Specifically, these volumes are computed by assuming water
molecules have van der Waals radii equal to $0.14$nm, and that the
diameter of each hydrophobic unit is $\sigma =0.72~$nm.. That is,
points in the excluded volume, ${\bf r}$, are those in the union of
all volumes inscribed by $|{\bf r}-{\bf r}_{\alpha }|<0.5$~nm, $\alpha
=1,2,...,N_{{\rm s}}$.}
\end{figure}

Brute-force simulations, confirmed by the analysis discussed below,
suggest that the collapse transition is a rare event. We therefore
performed transition-path-sampling simulations~\cite{Bolhuis02} to
harvest the rare, but representative, trajectories from the extended
coil to the collapsed globule state. Fig.~\ref{fig:snapshots} shows an
example of such a trajectory at room temperature. Initially, the
polymer is in the extended coil state. Then, by some spontaneous
fluctuation, the polymer partly collapses. The collapsed section forms
a sufficiently large hydrophobic cluster that a vapor bubble is
nucleated. Finally, the vapor bubble drives the beads of the polymer
together.

\newpage
\subsection{Free energy of collapse}
Visual inspections of this and similar 
trajectories suggest that the collapse arises from an interplay
between the size of the polymer and the formation of a bubble.

\begin{figure}[t]
\begin{center}
\epsfig{figure = 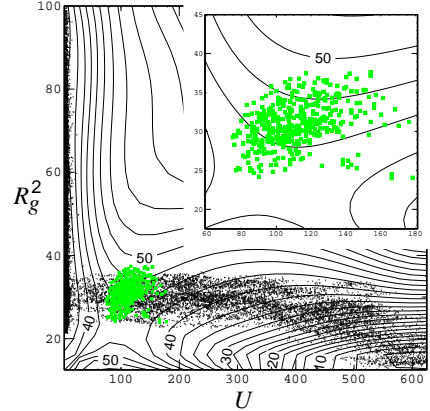,height=5.5cm}
\end{center}
\caption{\label{fig:umbr_tps}
Contour plot of the free energy landscape for the collapse of our
hydrophobic polymer, computed by Monte Carlo umbrella
sampling~\protect\cite {Frenkelbook}, using the weight functional
$\exp \left[ -\beta (V_{SS} ({\bf r}^{N_s}) + H ({\bf
r}^{N_s};\{n_k\})) \right]$. The curves of constant free energy are
drawn as a function of the squared radius of gyration of the polymer, $R_g^2$,
and the size of the largest bubble in the system, $U$. Here we have used
the ``cluster'' criterion that two vapor cells that are nearest neighbours,
belong to the same bubble; the size of the bubble is given by
the number of vapor cells. Neighboring lines of the free-energy
landscape are separated by
$2.5k_{B}T$. Superimposed is a scatter plot (in black) of the
harvested 150~ps trajectories going from the coil to the globule
state. The transition states are indicated in green. The harvesting
was performed with transition path sampling, making 8,400 moves in
trajectory space, of which 75\% were shooting and 25\% were
shifting~\protect\cite{Bolhuis02}. We find that the plateau regime of
the flux correlation function is reached after
50-70~ps~\protect\cite{Chandler78,Bolhuis02} implying that the typical
commitment time for trajectories to pass over the barrier is of the
order of 0.1ns. Given this time and the fact that the figure shows the
free energy barrier separating the extended coil and compact globule
states to be about $9k_{B}T,$ the half life of the extended chain is
about 0.1~ns $\, \times \exp \left( 9\right) \approx 10^{-5}$~s.
}
\end{figure}

 We therefore mapped out the free-energy landscape as a
function of the squared radius-of-gyration of the polymer, $R_g^2$,
and the size of the largest vapor bubble present in the system, $U$. A
contour plot of the free-energy landscape is shown in
Fig.~\ref{fig:umbr_tps}. It is seen that the path from the coil to the
collapsed globule is one where, initially, the radius-of-gyration
decreases, while the size of the largest vapor bubble is still
essentially zero. In this regime, the solvent still wets the polymer
and the free energy hardly changes. When the radius-of-gyration
becomes small enough, however, a vapor bubble is formed. The free
energy now sharply increases by some $9k_B T$, until it reaches a
saddle-point at $(U,R_g^2)=(98,23.5 l^2)$. From here on, the bubble
grows spontaneously and drives all the beads of the polymer together.

A close examination of the free-energy landscape reveals that there is
a small barrier of some $2 k_B T$ at $(U,R_g^2)=(6.5,70.0 l^2)$, which
separates the coil from a metastable intermediate. The presence of
these two states arises from a competition between the entropy of the
chain, which favors the fully extended coil state, and depletion
forces, which favor the intermediate state. Depletion forces are
caused by the reduction in the volume from which the solutes exclude
the solvent, when the solutes come together. The attraction between two
small hydrophobic objects, which cannot induce a drying transition in
the solvent, predominantly arises from this
effect~\cite{Pratt77,Pangali79,Hummer96}. However, the free-energy
landscape shows that this driving force is relatively small. The
large driving force for the collapse of the polymer comes from the
drying transition. The free-energy difference between the
fully-collapsed globule and
the intermediate is $30 k_BT$, whereas the free-energy difference
between the intermediate and the coil is only a few $k_B T$.

The collapse transition arises from an interplay between density
fluctuations of the solvent at small and large length scales. In the
coil state, the monomers are well separated and the solvation free
energy is dominated by the entropic cost of constraining small length
scale fluctuations. This cost scales with the size of the excluded
volume and, to a good approximation, is given by $\sum_i \Delta
\mu_{\rm ex}(v_i)$. But when the monomers come together, this entropic
cost is larger than the energetic cost of forming a vapor bubble that
envelopes the monomers. As water is close to phase coexistence, this
cost is dominated by the work to form an interface, which scales with
the area of the excluded volume. Indeed, it is the crossover in the
scaling behavior of the solvation free energy, as shown in
Fig.~\ref{fig:bdmu_SPCE}, which is the origin of the collapse
transition.

In order to make the above analysis more quantitative, we write the
free energy of ``folding'' as:
\begin{equation}
\label{eq:dG_fold}
\Delta G_{\rm fold} = \Delta G_{\rm solv} + \Delta G_{\rm intra}.
\end{equation}

\newpage

\begin{figure}[t]
\begin{center}
\epsfig{figure = 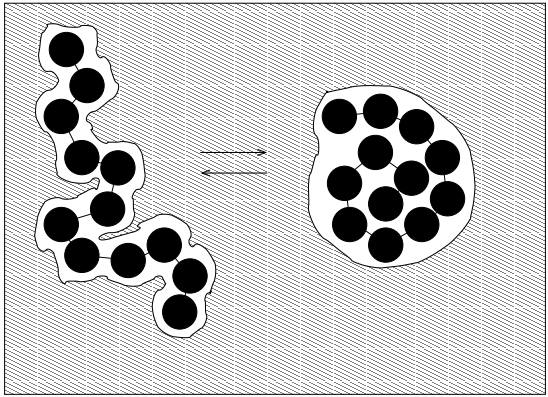,height=5.25cm}
\end{center}
\vspace*{0.25cm}
\caption{\label{fig:sketch_collapse}
Sketch of the collapse transition of the hydrophobic polymer. The
patterned region depicts the solvent, water. The white
region indicates the volume from which the polymer excludes solvent.
}
\end{figure}

Here $\Delta G_{\rm solv}$ is the reversible work to transfer the
chain from its ``wet'' solvated state to the ``dry'' state in the
vapor bubble, and $\Delta G_{\rm intra}$ denotes the free-energy change 
associated with the internal degrees of freedom of the polymer (see
Fig.~\ref{fig:sketch_collapse}). In order to estimate $\Delta G_{\rm
solv}$, we make the following assumptions: (1) the fully collapsed
globule is confined in a spherical bubble of radius $R$; (2) the
volume, from which the polymer excludes solvent in the extended coil
state is equal to the volume of the bubble that contains the globule;
(3) in the extended coil state, the solvent wets the polymer and
$n_i=1$ for all $i$. We then arrive at the following expression for
$\Delta G_{\rm solv}$:
\begin{eqnarray}
\Delta G_{\rm solv} &=& 4 \pi R^2 \gamma + \frac{4}{3} \pi R^3 p -
\sum_i \Delta \mu_{\rm ex} (v_i)\\
&=&4 \pi R^2 \gamma +  \frac{4}{3} \pi R^3 (p-c) \label{eq:dG_solv},
\end{eqnarray}
where $\gamma$ is the surface tension of the vapor-liquid interface,
$p$ is the external pressure and $c$ is given by
Eq.~\ref{eq:dmu_v}. Let us now compare the following numbers. First,
from the measured free-energy difference between the coil and the
globule, $\Delta G_{\rm fold}\approx -30 k_B T$, the radius of the
bubble, $R\approx 1.1$~nm, and Eqs.~\ref{eq:dG_fold}
and~\ref{eq:dG_solv}, we can obtain $\Delta G_{\rm intra}$. This
yields $\Delta G_{\rm intra} \approx 73 k_B T$. This should be
compared with the respective contributions to $\Delta G_{\rm solv}$,
$4\pi R^2 \gamma \approx 258 k_B T$ and $\frac{4 \pi R^3}{3} c \approx
361 k_B T$, which shows that the contribution of the internal degrees
of freedom of the polymer to the free energy of collapse is relatively
small. Indeed, the free energy of collapse of the hydrophobic polymer
is dominated by the solvation free energy. Further, at ambient
conditions $p<<c$, which shows that the collapse transition is
dominated by a competition between the energetic cost of creating a
vapor-liquid interface in the globule and the entropic cost of
maintaining the polymer in the wet state in the coil.

This simple model also allows us to address the
experimental observation that a wide range of proteins denature under
pressure. If we assume that (1) the specific volume of folding is
determined primarily by $\Delta G_{\rm solv}$ and that (2) the
dependence of the surface free energy upon pressure can be neglected,
then we arrive at:
\begin{equation}
\label{eq:dvfold}
\Delta v_{\rm fold} \approx \frac{\partial \Delta G_{\rm
solv}}{\partial p} \approx V (1- \frac{\partial c}{\partial p}).
\end{equation}
To our knowledge, no values for the dependence of the solvation free
energy upon pressure for hard spheres of the size used here, have
been reported. We therefore make some drastic assumptions. In order to
compute $\partial c/\partial p$ we assume that $c\approx k /\kappa_T$,
where $k$ is a constant, and $\kappa$ is the isothermal
compressibility. With $\kappa_T \approx 4.6 \times 10^{-10} {\rm
Pa}^{-1}$ at 1 atmosphere and $\kappa_T \approx 3.5\times 10^{-10}
{\rm Pa}^{-1}$ at
1000 atmosphere, we arrive at $\partial c/\partial p \approx
0.82$. This yields $\Delta v_{\rm fold} = V (1.-0.82) \approx 600 {\rm
cm}^3/{\rm mol}$. This is in agreement with the experimental
observation that proteins denature under pressure, although values
reported in the literature for proteins of similar size are
significantly smaller, by about a factor
six~\cite{Vidugiris95,Panick99}.

\subsection{Dynamics of collapse}
The free-energy landscape  shows that the rate-limiting step is
the formation of a vapor bubble of critical size. To investigate the
extent to which $R_g^2$ and $U$ also correctly describe the {\em dynamics}
of this nucleation event, we performed extensive
transition-path-sampling simulations on an ensemble of trajectories,
each of length 150 ps~\cite{TenWolde02_2}. A scatter plot of the
trajectories is shown in Fig.~\ref{fig:umbr_tps}. To identify the
transition state surface, we have identified the configurations on
each trajectory from which newly initiated trajectories have equal
probability of landing in either the coil or
globule~\cite{Bolhuis02,Northrup82}. These configurations are members
of the transition-state-ensemble. We project them onto
Fig.~\ref{fig:umbr_tps}. First, it is seen that the ensemble is
located close to the saddle-point in the free-energy landscape. It
should also be noted that the transition state ensemble is slightly
tilted in the $(U,R_g^2)$ plane. This shows that the larger the
polymer, the larger the size of the critical vapor bubble that has to
be nucleated. However, the scatter of the transition-state ensemble
from a line in the $(U,R_g^2)$ plane is notable, which indicates that at least
one other variable in addition to $U$ and $R_g^2$ plays a pertinent
role in the collapse transition.

It is also noteworthy that the transition paths do not follow the
lowest free-energy path from the metastable intermediate to the fully
collapsed globule (Fig.~\ref{fig:umbr_tps}). In particular, the
transition states are not centered around the saddle point and the
dynamical paths do not follow the steepest descend path from the
saddle point to the fully collapsed globule. The reason is that the
polymer and the solvent move on different time scales. The polymer
moves on time scales of nanoseconds, while the solvent moves on time
scales of picoseconds. As a result, when a vapor bubble is nucleated,
the chain cannot respond on the time scale at which the bubble is
formed. Only when the bubble has reached a size of around 400, does
the polymer fully collapse into the globule state.

The above observation indicates that free-energy
surfaces should be interpreted with care. The dynamical pathways for a
transition are not always fully determined by the underlying
free-energy surface. Fig.~\ref{fig:umbr_tps} shows that other aspects
can be important. In fact, when we artificially force the polymer to
move a hundred times faster with respect to the solvent (i.e. $M_s=3600$
instead of 36 -- see section~\ref{sec:Langevin}), then the transition
paths do follow the lowest free-energy path from coil to globule.
This clearly demonstrates that the dynamical trajectories are not only
determined by the underlying free-energy surface, but that the natural
dynamics of the system, or the dynamics of the algorithm, can be important
as well.  It also means that the transition
state ensemble does not need to coincide with the dividing surface in
the free-energy landscape, which is the surface that yields the
highest transmission coefficient~\cite{Bolhuis02}.

\section{Conclusions}
In this paper, we have given an overview of the solvation of
hydrophobic solutes. Its character is very different at small and
large length scales. At small length scales, solvation is dominated by
entropic effects and the solvent still wets the
surface of the solute, even when the solute is highly
hydrophobic~\cite{Lum99,Huang00}. In contrast, at large length scales,
solvation is dominated by enthalpic effects. In this regime, large
apolar species can induce a cavitation transition in the solvent. More
importantly, in the small length scale regime, the solvation free
energy scales with the volume of the solute, whereas in the drying
regime, the solvation free energy scales with the exposed area of the
solute. This behavior is not only observed for ideal hydrophobic
objects, but also for solutes that have weak dispersive interactions
with the solvent. The crossover behavior of the solvation free energy
from the wetting regime to the drying regime is important, because it
could be of significance to the formation of biological structures. In most
biological systems, the size of the hydrophobic species is such that
individual species are in the wetting regime, while assemblies of such
species are in the drying regime. In the wetting regime, water can
only induce a relatively weak attraction between two small apolar
species. When several of these species come together, however, water
can induce a strong attraction between them. 

A clear example of this process is given by the collapse of the
hydrophobic polymer. The strong driving force for the collapse of the
polymer, is provided by the cavitation transition in the
solvent. Importantly, the cavitation transition is only induced when a
sufficiently large number of monomers comes together. This means that
implicit solvent models, in which the interactions between the solutes
is described by a sum of two body terms, cannot describe the
cavitation transition. Equally important, the dynamics of the
hydrophobic collapse is dominated by the dynamics of the solvent,
especially the formation of a vapor bubble near the surface of a
nucleating cluster of hydrophobic monomers. As implicit solvent models
cannot capture the dynamics of the solvent, they are of limited use in
studying the effect of hydrophobic interactions on the {\em dynamics}
of biological self-assembly.

The connection between hydrophobic collapse and the cavitation/drying
transition provides a simple explanation for both cold
denaturation and pressure denaturation of proteins. The range and the
strength of the interactions between the hydrophobic objects in the
drying regime, is smaller when the solvent is moved away from phase
coexistence. Thus, the lowering of temperature and the increase of
pressure destabilize hydrophobic collapse, because both actions move
the solvent away from phase-coexistence. Finally, we believe that
fluorescence resonance energy transfer (FRET)
experiments~\cite{Deniz00} and nuclear magnetic resonance (NMR) spectroscopy~\cite{KleinSeetharaman02}
should make it possible to observe the drying transition in protein
folding experimentally.

\section*{Acknowledgments}
It is a pleasure to thank David Chandler and Paul Wessels for
carefully reading the manuscript.  This work has been supported in its
initial stages by the National Science Foundation (Grant No. 9508336
and 0078458) and in its final stages by the Director, Office of
Science, Office of Basic Energy Sciences, of the U.S. Department of
Energy (Grant No. DE-AC03-76SF00098).


\begin{thebibliography}{10}

\bibitem{Kauzmann59}
A.~Kauzmann, {\rm Adv. Prot. Chem.} {\bf 14}, 1 {(1959)}.\vspace{-0.3pt}

\bibitem{Tanfordbook}
C.~Tanford, {\em The Hydrophobic Effect - Formation of Micelles and Biological
  Membranes}, Wiley Interscience, New Yrok {(1973)}.\vspace{-0.3pt}

\bibitem{Pratt77}
{L. R. Pratt and D. Chandler}, {\rm J. Chem. Phys.} {\bf 67}, 3683
  {(1977)}.\vspace{-0.3pt}

\bibitem{Pangali79}
{C. Pangali, M. Rao, and B. J. Berne}, {\rm J. Chem. Phys.} {\bf 71}, 2975
  {(1979)}.\vspace{-0.3pt}

\bibitem{Chandler93}
{D. Chandler}, {\rm Phys. Rev. E} {\bf 48}, 2898 {(1993)}.\vspace{-0.3pt}

\bibitem{Hummer96}
{G. Hummer, S. Garde, A. E. Garc\'{i}a, A. Pohorille, and L. R. Pratt}, {\rm
  Proc. Natl. Acad. Sci. USA} {\bf 93}, 8951 {(1996)}.\vspace{-0.3pt}

\bibitem{Carambassis98}
{A. Carambassis, L. C. Jonker, P. Attard, M. W. Ruthland}, {\rm Phys. Rev.
  Lett.} {\bf 80}, 5357 {(1998)}.\vspace{-0.3pt}

\bibitem{Tyrrell01}
{J. W. G. Tyrrell and P. Attard}, {\rm Phys. Rev. Lett.} {\bf 87}, 176104
  {(2001)}.\vspace{-0.3pt}

\bibitem{Attard89}
{P. Attard}, {\rm J. Phys. Chem.} {\bf 93}, 6441 {(1989)}.\vspace{-0.3pt}

\bibitem{Eriksson89}
{J. C. Eriksson, S. Ljunggren, P. M. Claesson}, {\rm J. Chem. Soc. Faraday
  Trans. II} {\bf 85}, 163 {(1989)}.\vspace{-0.3pt}

\bibitem{Parker94}
{J. L. Parker, P. M. Claesson, and P. Attard}, {\rm J. Phys. Chem.} {\bf 98},
  8468 {(1994)}.\vspace{-0.3pt}

\bibitem{Attard00}
{P. Attard}, {\rm Langmuir} {\bf 16}, 4455 {(2000)}.\vspace{-0.3pt}

\bibitem{Evans90}
R.~Evans, {\rm J. Phys. Condens. Matter.} {\bf 2}, 8989
  {(1990)}.\vspace{-0.3pt}

\bibitem{Attard92}
{P. Attard, C. P. Ursenbach and G. N. Patey}, {\rm Phys. Rev. A} {\bf 45}, 7621
  {(1992)}.\vspace{-0.3pt}

\bibitem{Lum99}
{K. Lum, D. Chandler, and J. D. Weeks}, {\rm J. Phys. Chem. B} {\bf 103}, 4570
  {(1999)}.\vspace{-0.3pt}

\bibitem{Stillinger73}
F.~H. Stillinger, {\rm J. Solution Chem.} {\bf 2}, 141 {(1973)}.\vspace{-0.3pt}

\bibitem{Huang01_1}
{D. M. Huang, P. L. Geissler, and D. Chandler}, {\rm J. Phys. Chem. B} {\bf
  105}, 6704 {(2001)}.\vspace{-0.3pt}

\bibitem{TenWolde02_2}
{P. R. ten Wolde and D. Chandler}, {\rm Proc. Natl. Acad. Sci. USA}, to appear
  {(2002)}.\vspace{-0.3pt}

\bibitem{Huang00_1}
D.~M. Huang and D.~Chandler, {\rm Proc. Natl. Acad. Sci. USA} {\bf 97}, 8324
  {(2000)}.\vspace{-0.3pt}

\bibitem{TenWolde02_1}
{P. R. ten Wolde, S. X. Sun, and D. Chandler}, {\rm Phys. Rev. E} {\bf 65},
  011201 {(2002)}.\vspace{-0.3pt}

\bibitem{Crooks97}
G.~E. Crooks and D.~Chandler, {\rm Phys. Rev. E} {\bf 56}, 4217
  {(1997)}.\vspace{-0.3pt}

\bibitem{Huang00}
D.~M. Huang and D.~Chandler, {\rm Phys. Rev. E} {\bf 61}, 1501
  {(2000)}.\vspace{-0.3pt}

\bibitem{Hummer98}
{G. Hummer, S. Garde, A. E. Garc\'{i}a, M. E. Paulaitis, and L. R. Pratt}, {\rm
  J. Phys. Chem. B} {\bf 102}, 10469 {(1998)}.\vspace{-0.3pt}

\bibitem{Weeks98}
{J. D. Weeks, K. Katsov, and K. Vollmayr}, {\rm Phys. Rev. Lett.} {\bf 81},
  4400 {(1998)}.\vspace{-0.3pt}

\bibitem{Eisenberg86}
{D. Eisenberg and A. D. McLachlan}, {\rm Nature} {\bf 319}, 199
  {(1986)}.\vspace{-0.3pt}

\bibitem{Vallone98}
{B. Vallone, A. E. Miele, P. Vecchini, E. Chiancone, and M. Brunori}, {\rm
  Proc. Natl. Acad. Sci. USA} {\bf 95}, 6103 {(1998)}.\vspace{-0.3pt}

\bibitem{Car85}
{R. Car and M. Parinello}, {\rm Phys. Rev. Lett.} {\bf 55}, 2471
  {(1985)}.\vspace{-0.3pt}

\bibitem{Frenkelbook}
D.~Frenkel and B.~Smit, {\em Understanding Molecular Simulation, From
  Algorithms to Applications}, Academic Press, San Diego
  {(1996)}.\vspace{-0.3pt}

\bibitem{Lowen92}
{H. L\"{o}wen, P. A. Madden, and J.-P. Hansen}, {\rm Phys. Rev. Lett.} {\bf
  68}, 1081 {(1992)}.\vspace{-0.3pt}

\bibitem{Stillinger75}
F.~H. Stillinger, {\rm Adv. Chem. Phys.} {\bf 31}, 1 {(1975)}.\vspace{-0.3pt}

\bibitem{Landau00}
{D. P. Landau and K. Binder}, {\em A Guide to Monte Carlo Simulations
in Statistical Physics}, Cambridge University Press, Cambridge,
U.K. {(2000)}. \vspace{-0.3pt}

\bibitem{Huang02}
{D. M. Huang and D. Chandler}, {\rm J. Phys. Chem. B} {\bf 106}, 2047
  {(2002)}.\vspace{-0.3pt}

\bibitem{Guillot93}
{B. Guillot and Y. J. Guissani}, {\rm J. Chem. Phys.} {\bf 99}, 8075
  {(1993)}.\vspace{-0.3pt}

\bibitem{Bolhuis02}
{P. Bolhuis, D. Chandler, C. Dellago, and P.L. Geissler}, {\rm Ann. Rev. Phys.
  Chem.}, in press {(2002)}.\vspace{-0.3pt}

\bibitem{Vidugiris95}
{G. J. A. Vidugiris, J. L. Markley, and C. A. Royer}, {\rm Biochemistry} {\bf
  34}, 4909 {(1995)}.\vspace{-0.3pt}

\bibitem{Panick99}
{G. Panick G.J.A. Vidugiris, R. Malessa, G. Rapp, R. Winter, and C. A. Royer},
  {\rm Biochemistry} {\bf 38}, 4157 {(1999)}.\vspace{-0.3pt}

\bibitem{Northrup82}
{S. Northrup, M. R. Pear, C.-Y. Lee, A. McCammon, M. Karplus}, {\rm Proc. Natl.
  Acad. Sci. USA} {\bf 79}, 4035 {(1982)}.\vspace{-0.3pt}

\bibitem{Deniz00}
{A. A. Deniz, T. A. Laurence, G. S. Beligere, M. Dahan, A. B. Martin, D. S.
  Chemla, P. E. Dawson, P. G. Schultz, and S. Weiss}, {\rm Proc. Natl. Acad.
  Sci. USA} {\bf 97}, 5719 {(2000)}.\vspace{-0.3pt}

\bibitem{KleinSeetharaman02}
{J. Klein-Seetharaman, M. Oikawa, S. B. Grimshaw, J. Wirmer, E. Duchardt, T.
  Ueda, T. Imoto, L. J. Smith, C. M. Dobson, H. Schwalbe}, {\rm Science} {\bf
  295}, 1719 {(2002)}.\vspace{-0.3pt}

\bibitem{Chandler78}
D.~Chandler, {\rm J. Chem. Phys.} {\bf 68}, 2959 {(1978)}.\vspace{-0.3pt}

\end{thebibliography}
\end{document}